# Vortex-Beam Transient Absorption Microspectroscopy Resolves Ultrafast Free-Exciton and Polaron Diffusion in 2D Perovskites

Ju-Young Kim[1†], Anirban Mondal[1†] and Gi Rim Han[1]

[1]Center for Molecular Spectroscopy and Dynamics, Institute for Basic Science (IBS), Seoul 02841, Republic of Korea

[†]J.-Y.K. and A.M. contributed equally.

Kwang Jin Lee[2,3]

[2]Division of Semiconductor Physics, Korea University, Sejong 30019, Republic of Korea

[3]Department of Applied Physics, Graduate School, Korea University, Sejong 30019, Republic of Korea

Jong Min Lim[4] and Myeongsam Jen[4]

[4]Department of Chemistry and Green-Nano Materials Research Center, Kyungpook National University, Daegu 41566, Republic of Korea.

Minhaeng Cho[1,5*]

[5]Department of Chemistry, Korea University, Seoul 02841, Republic of Korea

* mcho@korea.ac.kr

## Abstract

Two-dimensional (2D) Ruddlesden–Popper perovskites are promising optoelectronic materials with strongly confined excitonic properties; however, probing their ultrafast carrier transport dynamics—particularly the initial nonequilibrium diffusion regime—remains challenging because conventional transient absorption microscopy requires complex spatial imaging and lacks sufficient temporal sensitivity to resolve early-time diffusion dynamics. Here, we demonstrate a vortex-beam-based transient absorption microspectroscopy platform (V-TAM) enabling imaging-free measurement of carrier transport by encoding spatial diffusion information into the mode-dependent pump–probe signal. By employing vortex probes with different topological charges, V-TAM provides mode-selective spatial sensitivity to excitonic dynamics with sub-picosecond temporal resolution. Using V-TAM, we resolved rapid free-exciton (FE) diffusion followed by relaxation toward a slower steady-state transport regime. A theoretically derived time-dependent diffusion model separated transient and steady-state transport contributions, yielding a transient diffusion enhancement (68.84 $cm^2/s$) and a steady-state diffusion coefficient (1.85 $cm^2/s$), thus providing an initial diffusion coefficient (70.69 $cm^2/s$), and a cooling time of 0.35 ps. Measurements at the exciton-polaron (EP) resonance revealed strongly suppressed diffusion with nearly time-independent signal ratios, indicating lattice-coupled EP transport. These parameters were extracted without spatial scanning or image reconstruction, establishing V-TAM as a powerful imaging-free platform for investigating carrier transport in perovskites and other semiconductor systems.

## 1. Introduction

Two-dimensional (2D) halide perovskites, particularly Ruddlesden–Popper (RP) structures, have recently emerged as a promising class of semiconductors due to their reduced dimensionality and highly tunable excitonic and optoelectronic properties.[1-4] These materials consist of inorganic octahedral layers separated by bulky organic spacer cations, forming a natural quantum well (QW) structure that enables strong dielectric confinement.[5] Compared to their three-dimensional counterparts, 2D RP perovskites exhibit enhanced environmental stability, bandgap tunability, and pronounced excitonic effects, characterized by increased exciton binding energies and reduced dielectric screening, making them promising for next-generation optoelectronic applications, including light-emitting diodes, lasers, and photovoltaics.[6-8] In these systems, strong exciton-phonon coupling leads to the formation of free excitons (FEs) and lattice-coupled exciton-polarons (EPs), which critically influence carrier relaxation and transport dynamics.[9] Despite extensive efforts to characterize excitonic properties, a comprehensive understanding of their dynamics remains elusive. In particular, the ability to disentangle and directly probe the diffusion of FEs and EPs is still limited, hindering a complete description of carrier transport mechanisms in 2D perovskites.

To investigate exciton transport in low-dimensional perovskites, several time-resolved optical microscopy techniques have been widely employed, among which time-resolved photoluminescence microscopy (TRPLM) and transient absorption microscopy (TAM) are the most commonly used.[10-13] Both techniques typically measure the variance of the exciton spatial distribution, $\sigma^2(t)$, as a function of time, rather than relying solely on temporal signals. The diffusion constant $D$ is then extracted using the relation $\sigma^2(t) = \sigma_0^2 + 2nDt$, where $n$ is the dimensionality of the diffusion (usually $n$ = 1 or 2 depending on the material).

TRPLM measures the spatial distribution of emissive excitons through photons emitted via radiative recombination, providing high sensitivity due to its background-free detection scheme. This approach offers intrinsic specificity to radiative states, which are directly relevant to light-emitting applications and solar cells. However, TRPLM is inherently limited to radiative (bright) excitons and is therefore blind to non-emissive species, including non-radiative excitons, triplet states, or trapped carriers. In addition, its temporal resolution is often limited by time-correlated single-photon counting (TCSPC) electronics to the tens of picoseconds regime, making it difficult to detect the early-stage ultrafast carrier transport. In contrast, TAM, a pump–probe-based technique, probes the photoinduced change in absorption ($\Delta A$) and allows tracking of the full population of photoexcited species, including both bright and dark excitons, polarons, and hot carriers.[14] Its temporal resolution, limited only by the laser pulse duration (typically 10–100 fs), provides access to ultrafast dynamics on the femtosecond timescale.[15] This capability enables the observation of nonequilibrium, or even ballistic, transport regimes prior to the onset of classical diffusion. By capturing both emissive and non-emissive states, TAM offers a more comprehensive picture of carrier dynamics that are not accessible through photoluminescence-based techniques (Table 1). Consequently, in the ultrafast time window of interest (0–1 ps), where transport deviates from classical diffusion, TAM is especially well suited—and among optical probes the most directly capable—for resolving early-stage carrier dynamics, owing to its femtosecond temporal resolution.

**Table 1. Comparison of TRPLM and TAM for investigating excitonic species in 2D perovskites.**

| Feature | TRPLM | TAM |
|---|---|---|
| **Species detected** | Radiative (bright) excitons | All (bright, dark, charges, etc.) |
| **Time resolution** | ~20–50 ps (TCSPC limited) | ~10–100 fs (pulse limited) |
| **Sensitivity** | Very high (background-free) | Moderate (differential signal) |
| **Primary limitation** | Missing dark (non-emissive state) dynamics | High experimental complexity |
| **Typical use case** | Long-range slow diffusion in organic films | Ultrafast transport in QDs or 2D materials[12, 14] |

Yet, despite its advantages, conventional TAM still suffers from several inherent limitations. As a differential measurement of a small photoinduced signal on a large optical background, TAM requires sophisticated modulation techniques, such as lock-in detection, and is more susceptible to laser noise than photoluminescence-based approaches. In particular, the signal at early delay times ($t \approx 0$) can be affected by pump–probe overlap, leading to coherent artifacts that require careful calibration and data interpretation. Furthermore, although TAM is the preferred technique for probing ultrafast dynamics in the 0–1 ps time window—where early-stage diffusion dynamics are established—it measures the total photoexcited population. As a result, it is necessary to disentangle population *decay* from spatial *expansion* (diffusion) when extracting diffusion dynamics from two-dimensional distributions.[13] This is particularly important in 2D perovskite QWs, where photoexcitation generates FEs (10–100 fs) that can undergo lattice dressing to form EPs (sub-ps timescale), as these species show distinct transport behaviors (Figure 1A). In conventional TAM, which employs Gaussian pump and probe beams, the reconstruction of spatial distributions requires the acquisition of time-resolved microscopy images, a process that is both time-consuming and experimentally demanding. Moreover, accurate extraction of the spatial variance relies on determining the center of the Gaussian profile in each frame, making the analysis highly sensitive to laser or stage drift and thereby compromising the reliability of the $\omega^2(t)$ evaluation (where $\omega$ denotes the beam waist).[16]

To overcome these limitations, we introduce a vortex-beam-based transient absorption microspectroscopy (V-TAM) technique that combines structured (vortex) light with TAM. A vortex beam is an optical field with a helical phase structure, characterized by a phase singularity at its center, giving rise to a doughnut-shaped intensity profile with a central intensity null.[17-19] The orbital angular momentum (OAM) carried by the vortex beam is quantified by its topological charge (TC, $l$); as the TC increases, the diameter of the doughnut profile, both its inner and outer radii, correspondingly expands (Figure 1B).[20-23]

The V-TAM platform utilizes a Gaussian pump and a vortex (doughnut-shaped) probe. In contrast to conventional Gaussian probes that yield spatially averaged signals, the vortex probe enables spatially selective detection, distinguishing different excitonic species by tuning the vortex beam radius to probe their evolving spatial distribution (Figure 1C, D). When excitons remain localized near the excitation center—within the central intensity null of the vortex probe (TC > 0)—the signal remains minimal due to low spatial overlap. As diffusion proceeds, the expanding exciton population spreads outward and increasingly overlaps with the doughnut-shaped probe's high-intensity regions, resulting in a time-dependent increase in the signal. This scheme functions as a spatial autocorrelation probe, significantly enhancing signal robustness and measurement efficiency by simplifying the detection architecture. By encoding spatial information directly into the beam geometry, V-TAM eliminates the need for complex, time-consuming imaging procedures or extensive spatial mapping typically required to resolve distinct excitonic species.

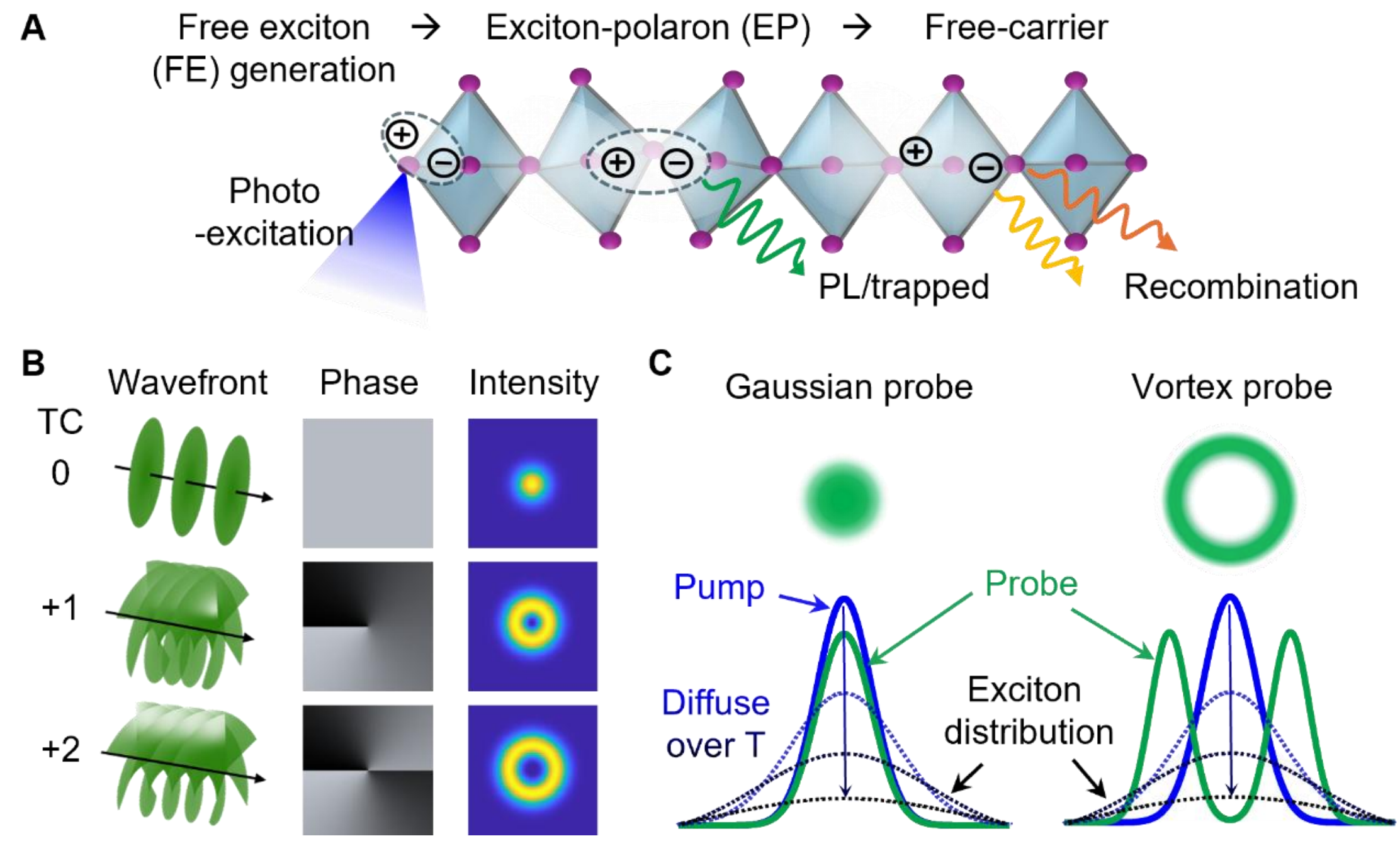

Fig. 1. Vortex beam-assisted TA measurements of exciton diffusion in 2D perovskites. (A) Diagram of exciton pathways in 2D perovskites. Photoexcited free excitons rapidly form exciton-polarons (10–100 fs) and predominantly decay via radiative recombination or trapping, with minimal dissociation into free carriers. (B) Helical wavefront, phase, and intensity profiles of vortex beams. (C) Exciton detection schemes using conventional Gaussian (left) and vortex (right) probe beams. Photoexcited exciton distribution broadens over time.

In this work, we implement V-TAM to probe ultrafast exciton transport in 2D RP perovskites at sub-picosecond time scales. Using $(PEA)_2PbI_4$ (PEA = phenethylammonium, $C_6H_5C_2H_4NH_3^+$) perovskite QWs, we provide a detailed investigation of early-stage dynamics, specifically resolving the transition from FEs to EPs. Supported by a robust theoretical framework, our results demonstrate that V-TAM directly captures the distinct transport signatures of these excitonic species that govern overall in-plane diffusion. By shifting from a scanning-based imaging requirement to a mode-overlap-based detection, we effectively transform a "microscopy problem" into a "spectroscopy solution". This approach allows a clear separation of population decay from spatial transport without time-consuming image reconstruction, establishing V-TAM as an efficient and powerful platform for investigating ultrafast transport phenomena in low-dimensional quantum materials.

## 2. Results

**Time-dependent diffusion equation.** As derived from our previous work,[24] the spatial diffusion contribution to the V-TAM signal, expressed as the ratio of the vortex ($l \neq 0$) to Gaussian ($l = 0$) probe signals, can be written as follows when the pump is a Gaussian, and the probe is a vortex beam with TC = $l_{\mathrm{pr}}$:

$$S(0, l_{\mathrm{pr}}; T) = \frac{2}{\pi} \frac{\left(w_{pu}^2 + 8DT\right)^{|l_{pr}|}}{\left(w_{pu}^2 + w_{pr}^2 + 8DT\right)^{|l_{pr}|+1}}. \quad (1)$$

Here, $D$ denotes the diffusion coefficient, $T$ is time, and $w$ represents the beam waist; $w_{pu}$ and $w_{pr}$ correspond to the pump and probe, respectively. The beam waist is given by $w = \lambda/(\pi \cdot NA)$, where $\lambda$ is the wavelength and the $NA$ is the numerical aperture of the excitation lens. Parameters used in our work are as follows: $\lambda_{pu} = 400$ nm, $\lambda_{pr} = 520/530$ nm, and $NA = 0.65$. Assuming a diffraction-limited Gaussian 1/e$^2$ waist at focus, $w_0 \approx {}^{\lambda}/_{\pi\,\mathrm{NA}}$. Accordingly, the Gaussian waists at focus are $w_{pu} = 0.196\ \mu m$ and $w_{pr} = 0.255\ \mu m.$ Since the vortex probe is an $LG_{0,l}$ mode, its intensity maximum lies on a bright ring with diameter $d_l = 2r_l = w_{pr}\sqrt{|l|/2}$. Therefore, the focused doughnut diameters are 0.36 μm for TC = 1 and 0.51 μm for TC = 2. Here, $w_{pr}$ denotes the Gaussian waist parameter of the vortex mode used in Eq. (1), not the doughnut diameter.

Notably, for the special case of $l_{pr} = 0$ (a Gaussian probe), Eq. (1) reduces to the well-known signal expression employed in fluorescence (intensity) correlation spectroscopy (FCS) for a Gaussian focal volume,[25] in which the temporal decay of the signal directly reflects the diffusion of fluorescent species through a Gaussian observation volume. Eq. (1) generalizes this expression to the case of a vortex probe ($l_{pr} \neq 0$), where the time-dependent spatial overlap between the Gaussian pump and the donut-shaped vortex probe is exploited to encode the diffusion-driven population spread into the temporal evolution of the mode-resolved V-TAM signal.

Starting from this expression for a constant diffusion coefficient, we now extend the model to account for hot-carrier transport by introducing a time-dependent diffusion coefficient $D(t)$. Hot carriers with high kinetic energy, generated by a pump laser tuned well above the bandgap, can exhibit substantially enhanced diffusion compared with thermalized carriers; the diffusion coefficient must therefore be treated as time-dependent, $D(t)$, evolving as the carriers lose kinetic energy and relax toward the band edge. The corresponding diffusion equation $\partial n/\partial t = D(t)\nabla^2 n$ can be solved analytically by reducing it to the standard diffusion form through a change of the time variable (full derivation in Supplementary Note 1).

We model the instantaneous diffusion rate as $D(t) = D_{bulk} + \delta D \cdot e^{-t/t_{cool}}$, where $t_{cool}$ represents the characteristic cooling time of hot carriers relaxing toward the band edge. Accordingly, the diffusion rate starts from a large diffusion coefficient of $D_{bulk} + \delta D$ at $t = 0$ and decays exponentially to the steady-state bulk value $D_{bulk}$. The relevant quantity entering the V-TAM signal is the accumulated diffusion over the pump–probe delay $T$, obtained by integrating $D(t)$:

$$\Lambda(T) = \int_0^T D(t)dt = D_{bulk} \cdot T + \delta D \cdot t_{cool} \cdot \left(1 - e^{-T/t_{cool}}\right) \quad (2)$$

This quantity directly governs the time-dependent broadening of the carrier distribution and is proportional to the mean-squared displacement, $\langle r^2 \rangle = 2m \cdot \Lambda(T)$, where $m$ is the dimensionality of diffusion. Initially, the large excess kinetic energy of hot carriers drives a high diffusion flux; as carriers couple to the lattice via phonon emission, this driving force vanishes, and the system relaxes into the bulk diffusion regime.

The spatial diffusion contribution to the V-TAM signal under time-dependent diffusion is then obtained by replacing $D \cdot T$ in Eq. (1) with $\Lambda(T)$ (Supplementary Note 1). The complete V-TAM signal further includes a mode-dependent amplitude prefactor $a_{l_{pr}}$ (accounting for differences in intensity normalization and detection efficiency across vortex modes) and a population-decay term $e^{-T/t_{ex}}$ describing exciton relaxation; the full expression is given in Supplementary Note 1 (Eq. (S16)). To eliminate the dependence on amplitude calibration and on the excitation lifetime, we analyze the signal ratio with respect to the Gaussian (TC = 0) probe,

$$R_{l_{pr}0}\left(l_{pr}, T\right) \equiv S_{V-TAM}\left(l_{pr}, T\right) / S_{V-TAM}\left(l_{pr} = 0, T\right), \quad (3)$$

in which the population-decay term $e^{-T/t_{ex}}$ cancels exactly between numerator and denominator. Specifically, we denote $R_{10}(T) \equiv S_{V-TAM}\left(l_{pr} = 1, T\right) / S_{V-TAM}\left(l_{pr} = 0, T\right)$ and $R_{20}(T) \equiv S_{V-TAM}\left(l_{pr} = 2, T\right) / S_{V-TAM}\left(l_{pr} = 0, T\right)$ for the TC of 1 and 2 vortex probes, respectively. As a result, the extracted diffusion parameters are independent of the exciton lifetime, even for large $t_{ex}$ (e.g., $t_{ex}$ >100 ps), and the analysis is determined entirely by the diffusion-driven temporal evolution encoded in $\Lambda(T)$.

**2D perovskites.** $(PEA)_2PbI_4$ were synthesized as a quantum-well structure (see Materials and methods section), where inorganic slabs are confined between organic spacer layers (Figure 2A). The morphology of the as-prepared films was characterized by scanning electron microscopy (SEM) and atomic force microscopy (AFM), as shown in Figure S1. The relevant photoexcitation pathways are illustrated in Figure 2B, where hot carriers relax to form FEs, which subsequently undergo lattice dressing to form EPs before decaying via radiative recombination or trapping/dissociation. The formation of these excitonic species was examined using steady-state absorption and photoluminescence (PL) spectroscopy for $(PEA)_2PbI_4$ (Figure 2C). Under 400-nm excitation, the sample exhibits strong PL emission centered at 535 nm. Gaussian decomposition of the PL spectrum identifies two spectral components tentatively attributed to FE and EP contributions, consistent with prior assignments in $(PEA)_2PbI_4$.[1, 8, 26]

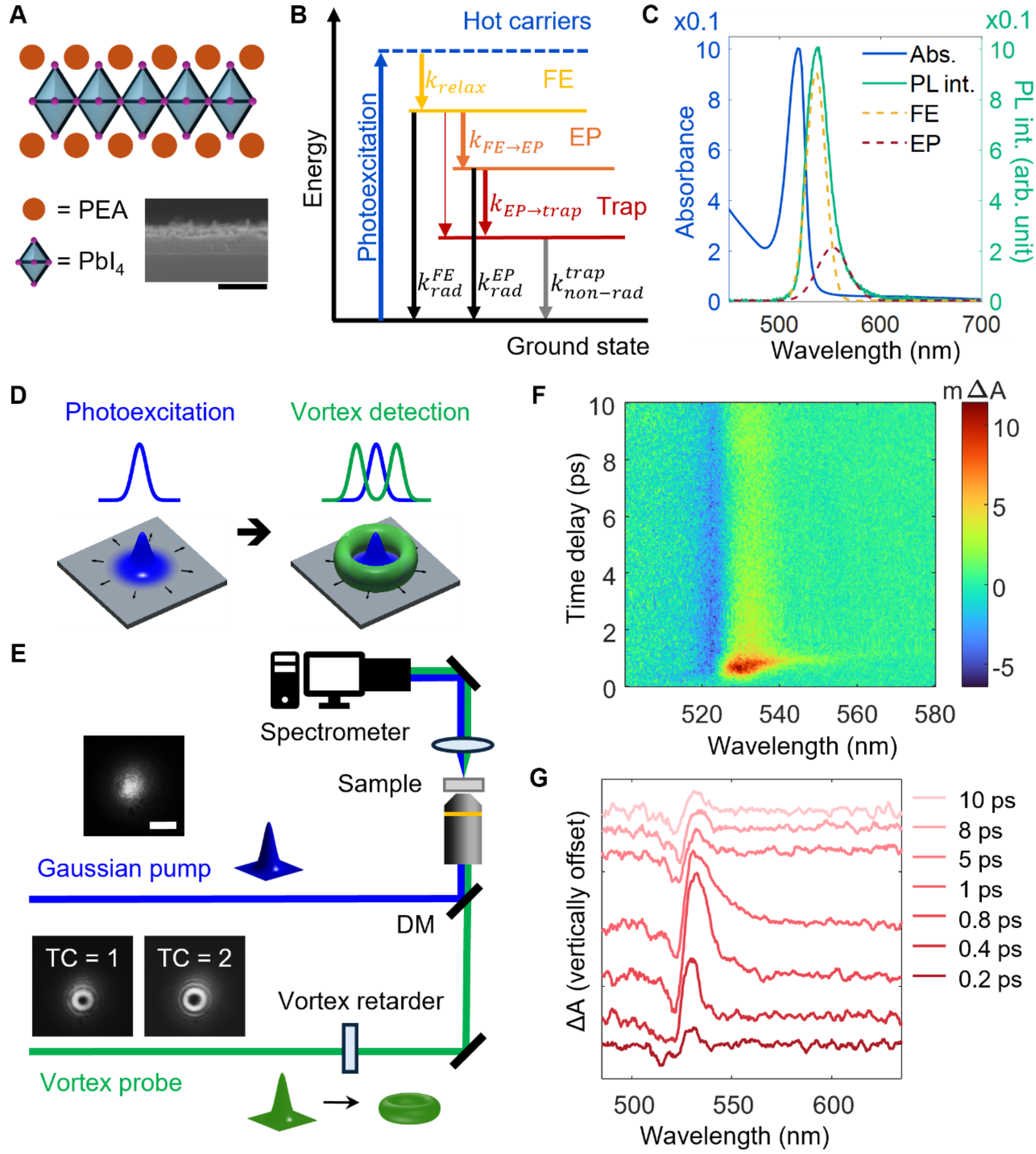

**Fig. 2. Structural and optical characterization of 2D perovskites.** (A) Crystal structures of $(PEA)_2PbI_4$. Inset: SEM image of a perovskite layer, with a scale bar of 500 nm. (B) Energy diagram of exciton species. Radiative decay pathways ($k_{rad}$) from these states give rise to PL. (C) Steady-state absorption and PL spectra of $(PEA)_2PbI_4$. Gaussian convolutions show averaged exciton species, FEs and EPs. (D) V-TAM measurements: a Gaussian pump generates a localized exciton population that diffuses out over time (left), and a vortex probe monitors its spatial evolution (right). (E) Schematic of the V-TAM setup. Vortex probe beams are generated by a vortex retarder and combined with a Gaussian pump via a dichroic mirror (DM). After interaction with the perovskite sample, the V-TAM signal is detected using a spectrometer. Insets show the intensity profiles of the Gaussian pump and vortex probes (scale bar, 10 μm). (F) Delay time-dependent TA maps and (G) representative wavelength traces extracted from the TA maps.

**V-TAM measurements.** To elucidate the spatial transport dynamics of the identified excitonic species, we employed a vortex beam as a probe of the TA measurements. As illustrated in Figure 2D, photoexcited excitons undergo spatial diffusion and subsequent lattice dressing. A doughnut-shaped vortex beam is used to selectively monitor the spatial spread of the exciton population. Figure 2E provides the schematic of our custom-built V-TAM platform based on imaging-free vortex probe measurements. A vortex retarder converts a Gaussian probe into a vortex beam by introducing helical phase profile. The vortex probe diameter is controlled by the TC of the retarder, such that TC = 1 and TC = 2 correspond to increasingly larger spatial regions. Consequently, the V-TAM signal reflects the time-dependent spatial overlap between the diffusing exciton population and the vortex probe, providing direct sensitivity to exciton transport dynamics. A detailed description of the optical setup is provided in the Experimental Section.

To identify the spectral signatures of these excitonic species, conventional TA measurements using a Gaussian probe (TC = 0) were first performed for both perovskites (Figure 2F), revealing distinct spectral features corresponding to FEs (520 $\pm$ 5 nm) and EPs (530 $\pm$ 3 nm). The temporal evolution of these features is further illustrated in Figure 2G, where representative wavelength traces highlight their distinct dynamics. The identified FE and EP resonances were subsequently used as probe wavelengths in the V-TAM measurements, allowing the transport dynamics of each excitonic species to be selectively monitored.

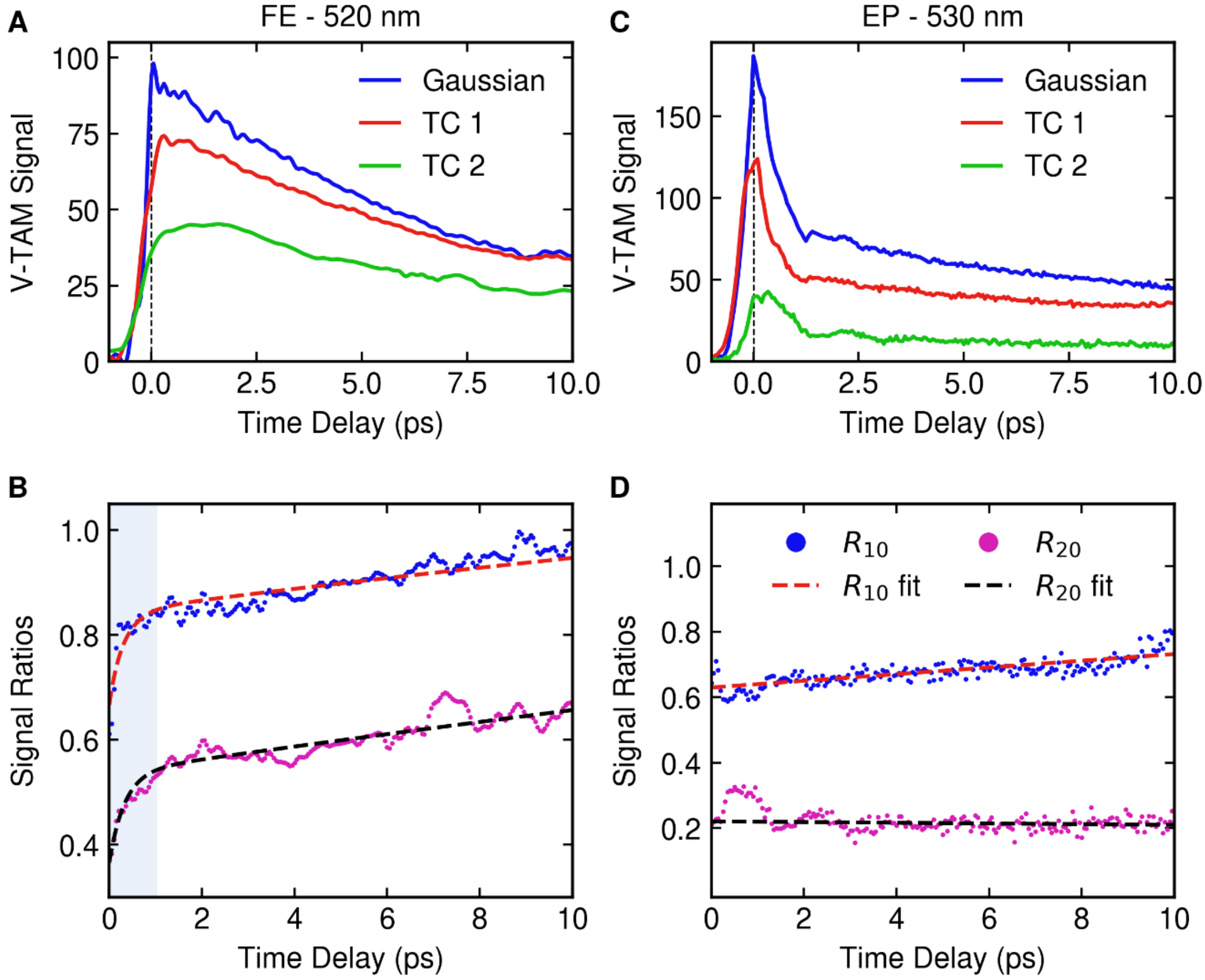

**Figure 3. Time-dependent diffusion fitting of the V-TAM signal ratios.** (A) Experimentally measured V-TAM spectra using Gaussian and vortex probes with TC of 1 and 2 at the 520 nm free-exciton resonance. (B) Signal ratios, $R_{10}$ and $R_{20}$, with the corresponding fits using a time-dependent diffusion model. (C) V-TAM spectra measured at the 530 nm exciton-polaron resonance. Expanded view of the early-state diffusion regime, highlighting the rapid initial increase in the signal ratios associated with hot-carrier diffusion. (D) Signal ratios, $R_{10}$ and $R_{20}$, with the corresponding linear fits.

**Analysis of transport dynamics.** Using the FE and EP resonance wavelengths identified above, V-TAM measurements were performed with Gaussian and vortex probes to track the spatial evolution of each excitonic species. To quantitatively analyze the time-dependent excitonic transport dynamics, we applied the time-dependent diffusion model to the experimentally measured signal ratios. Figure 3A shows the deconvolved V-TAM spectra measured using Gaussian and vortex probes with TC of 1 and 2. The deconvolution procedure using the instrument response function is described in Figure S2. For the transport analysis, the time origin ($T = 0$) was set after the initial signal build-up, when the carrier distribution is spatially established, and the subsequent dynamics are dominated by diffusion. Periodic oscillations originating from coherent phonon modes of the $(PEA)_2PbI_4$ lattice are superimposed on the signal curves and were isolated via Fourier-based filtering. The three dominant frequencies extracted from this procedure — 70, 83, and 103 $cm^{-1}$, assigned to Pb–I bending (scissoring) and stretching — are in good agreement with literature values reported for Pb–I vibrations, confirming their lattice-vibrational origin. Details of the FFT filtering process and mode assignment are provided in Figure S3.

Following Eq. (3), the signal ratios, $R_{10}$ and $R_{20}$, with the corresponding fits using a time-dependent diffusion model are presented in Figure 3B. A rapid rise in both ratios is shown in the early-time regime, reflecting the hot-carrier-enhanced diffusion that decays as the carriers cool toward the steady-state regime. The experimental signal ratios were fitted using the full time-dependent diffusion model, which resolves the transient evolution of the diffusion dynamics from the initial nonequilibrium regime to the late-time steady-state regime and enables the explicit extraction of the key transport parameters: $\delta D$, $D_{bulk}$, and $t_{cool}$. Note that the two signal ratios $R_{10}$ and $R_{20}$ were fitted with a single shared set of parameters (Supplementary Note 1). The optimized fitting results reveal a cooling time, $t_{cool} \sim 0.35$ ps, which accounts for the rapid initial curvature observed within the first picosecond, reflecting the relaxation of the initially nonequilibrium carrier population. This analysis further indicates an enhanced initial diffusion contribution of $\delta D \sim 68.84$ cm$^2$/s, corresponding to substantially faster carrier transport during the early nonequilibrium regime. After the hot-carrier cooling process, the transport dynamics evolve toward a steady-state diffusion regime governed by a bulk diffusion coefficient of $D_{bulk} \sim 1.85$ cm$^2$/s, as determined from the recovery phase between 2 and 10 ps. See Table S1 for the list of fitting parameters.

We then extended our V-TAM measurements to the EP resonance at 530 nm using vortex beams of TC = 1 and 2, as shown in Figure 3C. The corresponding deconvolution and Fourier filtering procedure using the instrument response function are shown in Figures S4 and S5. Following the same analysis procedure described above, the experimentally measured signal ratios $R_{10}$ and $R_{20}$ were extracted (Figure 3D). Linear fitting of the signal ratios provided nearly zero slope ($m_R = 0.01$ ps$^{-1}$ and $m_F = -0.001$ ps$^{-1}$), respectively. In contrast to the conspicuous time-dependent evolution observed at the 520 nm measurements, the nearly time-independent behavior of the 530 nm signal ratios indicates strongly suppressed diffusion dynamics. This behavior is consistent with the formation of EP, in which lattice-coupled quasiparticles with enhanced effective masses exhibit substantially reduced diffusion relative to FE transport.

## 3. **Discussion and conclusion**

The V-TAM measurements, combined with structured-light (vortex) signal-ratio analysis, provided direct access to ultrafast excitonic transport dynamics in the $(PEA)_2PbI_4$ 2D perovskite. Direct fitting of $R_{10}$ and $R_{20}$ with the theoretically derived time-dependent diffusion model resolves the distinct nonequilibrium and steady-state regimes, yielding an enhanced initial diffusion contribution of $\delta D$ ~68.84 $cm^2/s$ with a cooling time of $t_{\mathrm{cool}}$ ~0.35 ps, alongside a late-time steady-state coefficient of $D_{\mathrm{bulk}}$ ~1.85 $cm^2/s$. Therefore, the total initial diffusion coefficient amounts to $D(0) = D_{bulk} + \delta D$ ~70.69 $cm^2/s$.

These results are consistent with previously reported ultrafast TAM studies, in which hot carriers exhibit substantially enhanced diffusion during the first few hundred femtoseconds after photoexcitation, before cooling into slower, steady-state regimes. First, $D_{\mathrm{bulk}}$ falls squarely within the range of fast FE transport regimes reported for high-quality 2D perovskites, well above the slow baseline diffusion (< 1 $cm^2/s$) typically observed in conventional RP films.[27, 28] Recent studies on surface-engineered and highly crystalline 2D perovskite flakes have reported fast-diffusion coefficients of 1.2–1.9 $cm^2/s$,[11, 29, 30] consistent in magnitude with our result. As for the initial hot-carrier enhancement, the relatively lower $\delta D$, compared with 3D perovskites and graphene systems that show exceptionally large hot-carrier coefficients,[13, 31, 32] is physically reasonable, given the stronger exciton confinement and enhanced carrier-phonon interactions of the 2D quantum-well perovskite structure. A comparison between the constant and transient diffusion coefficients obtained in this work and those reported in previous ultrafast transport studies is summarized in Table 2. Importantly, unlike conventional ultrafast diffusion measurements that rely on spatial mapping or microscopy-based imaging, the present V-TAM approach extracts transient diffusion dynamics directly from spectroscopic mode-overlap analysis alone.

**Table 2. Comparison between measured and reported excitonic diffusion coefficients.**

| Ref | Materials | D ($cm^2/s$) | Method |
| --- | --- | --- | --- |

| **[13]** | Guo et al.,<br>Science 2017 | 3D perovskite: $MAPbI_3$ | 450±10 (hot 1 ps) → 0.7 (eq.) | Ultrafast TA microscopy* |
|---|---|---|---|---|
| **[31]** | Tan et al.,<br>APL 2024 | 3D perovskite: $CsSnBr_3$ | 257.8±1.8 (initial)<br>→12.7-16.5 (cooled) | |
| **[32]** | Ruzicka et al.<br>Phys Rev B 2010 | Graphene | 5500-11000 (sub-ps hot) | |
| **[29]** | Xiao et al.,<br>Adv Mater 2020 | 2D perovskite:<br>$(mF\text{-}PEA)_2PbBr_4$ | 1.91 | TRPLM |
| **[30]** | Terres et al.,<br>Adv Opt Mater 2025 | 2D perovskite:<br>$(MBA)_2PbI_4$ | 1.2 -1.4 | |
| **[11]** | Gong et al.,<br>Nat Comm 2024 | 2D perovskite:<br>$(BA)_2(MA)_{n-1}Pb_nI3_{n+1}$ | 7.65 (fast) →1.88 (slow) | |
| ***Our work*** | | 2D perovskite:<br>$(PEA)_2PbI_4$ | 70.69 (initial, = $D_{\text{bulk}} + \delta D$ ;<br>$\delta D$ = 68.84) → 1.85 ($D_{\text{bulk}}$) | ***V-TAM:<br>no need imaging*** |

* Highlighted are reported $\delta D$ values, while others are $D_{\text{bulk}}$.

Physically, the substantially enhanced initial diffusion coefficients, $D_{initial}$ ($= D_{bulk} + \delta D \approx$ 70.69 cm$^2$/s), can be rationalized by two complementary physical arguments. First, according to the Einstein relation, $D = \mu k_B T_c / e$, the diffusion coefficient scales with the carrier temperature ($T_c$). Immediately after photoexcitation, $T_c$ can transiently reach several thousand Kelvin, producing a large transient enhancement in $D$. Taking $D_{\text{bulk}}$ to represent the room-temperature regime, the observed ratio $D_{hot}/D_{bulk}$ ~38 would imply an effective initial $T_c$ on the order of $10^4$ K if attributed to temperature alone. However, an independent estimate of $T_c$ from the excess energy of the pump gives a lower value: with a 3.10 eV pump and the $(PEA)_2PbI_4$ FE resonance at ~2.38 eV, the per-carrier excess energy is $\sim 0.36$ eV, corresponding to $T_c$~2800–4200 K depending on whether 3D or in-plane 2D equipartition is assumed (Supplementary Note 2 provides the detailed calculation). This indicates that the elevated carrier temperature accounts for only part of the enhancement. Second, the remainder is attributed to a transient increase in carrier mobility $\mu$: at elevated carrier energies, hot carriers are less susceptible to acoustic-phonon and ionized-impurity scattering, raising the effective mobility $\mu$ ($D \propto \mu T_c$). Together, these arguments strongly support the interpretation that the rapid early-time evolution observed in V-TAM originated from transient hot-carrier excitonic transport prior to thermal equilibrium. The estimated $T_c$ values and corresponding ultrafast-transport studies are summarized in Table 3.

**Table 3. Comparison of experimental conditions and reported hot-carrier diffusion coefficients across representative ultrafast TAM studies.**

| Ref | Materials | Pump E | Probe E | Bandgap | Excess E (eV) | Tc (K) | $D_{initial}$ ($cm^2/s$) |
|---|---|---|---|---|---|---|---|
| **[13]** | MAPbI3 | 3.14 eV | 1.58 eV | - | 1.49 eV | - | 450 $\pm$ 10 |
| **[31]** | CsSnBr3 | 515 nm | 650 nm | 1.75 eV | 0.66 eV* | ~1,500 | 257.8 $\pm$ 1.8 |
| **[32]** | Graphene | 750 nm | 810 nm | (gapless) | > 0.8 eV | ~3,600 | 11,000 |
| ***Our work*** | $(PEA)_2PbI_4$ | 400 nm | 520 nm | 2.38 eV | 0.71 | ~$10^4$ | 70.69 |

* derived from the values

In addition to probing the highly mobile FE regime at 520 nm, V-TAM measurements at the 530 nm resonance, associated with the EP regime, exhibited nearly time-independent signal ratios with near-zero slopes (Figure 3), indicating strongly suppressed diffusion dynamics. This behavior is consistent with EP formation, in which excitonic quasiparticles are dressed by lattice distortions in the highly polarizable Pb–X framework, leading to an enhanced effective mass and reduced diffusion. The relatively stable but weak transport observed at 530 nm further suggests lattice-coupled excitonic transport characteristic of low-dimensional (2D) perovskites with strong electron–phonon coupling. Collectively, these results demonstrate that V-TAM successfully resolves the highly mobile nonequilibrium FE regime from the stabilized EP regime—without spatial imaging or scanning-based microscopy.

Because the FE and EP probe windows at 520 and 530 nm are spectrally close (10 nm bandwidth), partial spectral overlap between the two species may contribute to the measured signals. In addition, the extracted diffusion coefficient is sensitive to the probe spectral range, as wavelength-integrated analysis mixes distinct carrier populations and yields substantially different diffusion values compared with the single-wavelength probing used in this work (Supplementary Note 3). Nevertheless, the pronounced contrast between the strongly time-dependent FE signals and nearly flat EP response confirms that the two transport regimes remain well resolved under the band-pass-filtered V-TAM measurement.

In conclusion, we have proposed and experimentally demonstrated the V-TAM platform as a powerful approach for probing ultrafast excitonic transport dynamics in low-dimensional perovskites. By combining structured-light mode-overlap spectroscopy with time-dependent diffusion analysis, V-TAM directly accesses the transient branching between highly mobile nonequilibrium FE transport and stabilized EP states with sub-picosecond temporal sensitivity. Unlike conventional ultrafast spatial microscopy techniques, the present approach extracts diffusion dynamics without requiring spatial imaging or scanning-based reconstruction. These results establish V-TAM as a promising platform for probing nonequilibrium carrier transport and quasiparticle formation in $(PEA)_2PbI_4$, with broad applicability to other semiconductor systems pending validation.

## 4. Experimental Section

**2D perovskite thin films synthesis.** Lead iodide ($PbI_2$), phenylethyl ammonium iodide (PEAI), and N, N-dimethylformamide (DMF) were purchased from Sigma-Aldrich and used as received. To prepare the $(PEA)_2PbI_4$ precursor solution, PEAI and $PbI_2$ were dissolved in 1 mL of DMF (0.8 and 0.4 M, respectively), followed by magnetic stirring for 30 min to obtain a homogeneous solution. Quartz substrates were cleaned by sequential sonicating in detergent, acetone, and isopropyl alcohol, followed by a 20-min oxygen plasma treatment. The precursor solution was then spin-coated onto the plasma-treated substrates at 3000 rpm for 30 s, followed by annealing at 100 °C for 10 min.

**Optical Absorption and Photoluminescence Spectroscopy.** UV–Vis absorption and photoluminescence (PL) spectra of the 2D perovskite thin-films were acquired at room temperature using a Duetta system (Horiba Scientific) equipped with a xenon lamp as the excitation source.

**V-TAM optical setup.** A custom-built V-TAM system was employed to probe ultrafast exciton dynamics. To generate the vortex probe beam, the Gaussian probe was passed through an optical vortex retarder (Thorlabs, TC = 1 or 2) before being focused onto the sample. Single-color probe wavelengths centered at 520 and 530 nm were used to selectively monitor free-exciton and exciton-polaron dynamics, respectively. The experimental setup and measurement procedures are schematically illustrated and described in Figure S6.

**Acknowledgements**

This work was supported by the Institute for Basic Science (IBS) under Grant No. IBS-R023-D1. SEM and AFM image analyses were carried out at the Korea Advanced Nano Fab Center (KANC).

# Supplementary Information for

# Vortex-Beam Transient Absorption Microspectroscopy Resolves Ultrafast Free-Exciton and Polaron Diffusion in 2D Perovskites

*Ju-Young Kim[1†], Anirban Mondal[1†], Gi Rim Han[1], Kwang Jin Lee[2,3], Jong Min Lim[4], Myeongsam Jen[4], and Minhaeng Cho[1,5*]*

[1]Center for Molecular Spectroscopy and Dynamics, Institute for Basic Science (IBS), Seoul 02841, Republic of Korea.
[2]Division of Semiconductor Physics, Korea University, Sejong 30019, Republic of Korea
[3]Department of Applied Physics, Graduate School, Korea University, Sejong 30019, Republic of Korea
[4]Department of Chemistry and Green-Nano Materials Research Center, Kyungpook National University, Daegu 41566, Republic of Korea.
[5]Department of Chemistry, Korea University, Seoul 02841, Republic of Korea.
* mcho@korea.ac.kr

[†]J.-Y. K. and A.M. contributed equally.

**Table of contents**

**Figure S1. SEM and AFM images of the perovskite film**

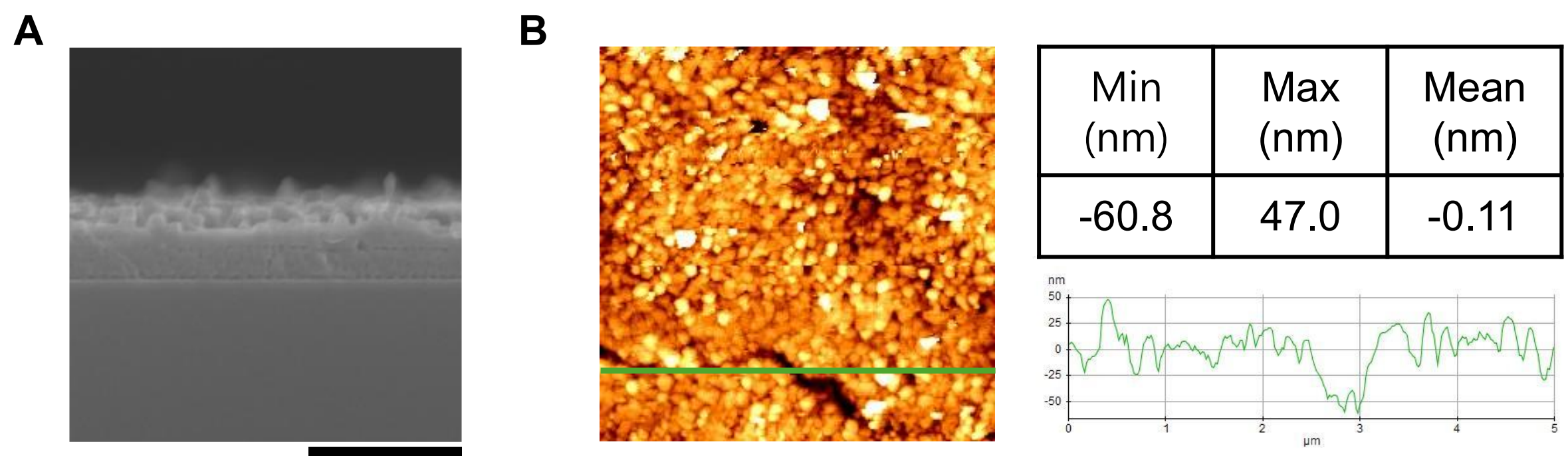

| Min (nm) | Max (nm) | Mean (nm) |
|---|---|---|
| -60.8 | 47.0 | -0.11 |

(A) Cross-sectional SEM image of the $(PEA)_2PbI_4$ film showing PEA spacer-induced upper surface morphology with uniform film thickness. Scale bar: 1 µm. (B) AFM image and corresponding height profile of the perovskite. The film thickness was estimated to be approximately 59.7 nm from the height difference between the maximum and minimum values. Scan area: 5×5 $\mu m^2$ with 256×256 pixels.

**Supplementary Note 1. Solution of the time-dependent diffusion model**

In the main text, the spatial diffusion contribution to the V-TAM signal is modeled using a time-dependent diffusion coefficient $D(t)$, and the accumulated diffusion $\Lambda(T)$ is introduced as the quantity that enters the signal expression. Here, we provide the complete mathematical derivation of how the time-dependent diffusion equation reduces to the standard diffusion form, and derive the explicit Gaussian solutions used to relate $\Lambda(T)$ to the V-TAM signal, and write out the full V-TAM signal expression used in the fitting.

1) Reduce it to the standard diffusion equation with a time change

Hot carriers with excess kinetic energy can exhibit transiently enhanced diffusion before cooling toward the band edge, motivating the use of a time-dependent diffusion coefficient $D(t)$.[1] As such, the governing partial differential equation for the carrier density is given by

$$\frac{\partial n(\mathbf{r},t)}{\partial t} = D(t)\,\nabla^2 n(\mathbf{r},t) \tag{S1}$$

This time-dependent equation can be solved as long as $D(t)$ depends only on time $t$ (not on $\mathbf{r}$). We define the integrated (accumulated) diffusivity:

$$\Lambda(t) \equiv \int_0^t D(s)\,ds \;(\Lambda \geq 0 \text{ if } D \geq 0). \tag{S2}$$

Treat $n(\mathbf{r},t) = n(\mathbf{r},\Lambda(t))$. By the chain rule,

$$\frac{\partial n}{\partial t} = \frac{d\Lambda}{dt}\frac{\partial n}{\partial \Lambda} = D(t)\frac{\partial n}{\partial \Lambda}.$$

Substituting into the PDE,

$$D(t)\frac{\partial n}{\partial \Lambda} = D(t)\nabla^2 n \Rightarrow \frac{\partial n}{\partial \Lambda} = \nabla^2 n. \tag{S3}$$

This reduces the time-dependent problem to the standard heat/diffusion equation with a unit diffusion coefficient, in which the effective time variable is $\Lambda$. The variable $\Lambda$ corresponds directly to the accumulated diffusion introduced in the main text, Eq. (2).

2) Solution in Fourier space (works in any dimension)

Taking a spatial Fourier transform:

$$\hat{n}(\mathbf{k},t) = \int d^d\mathbf{r}\, e^{-i\mathbf{k}\cdot\mathbf{r}} n(\mathbf{r},t). \tag{S4}$$

where superscript $d$ denotes the spatial dimensionality of the system, so that $d^d\mathbf{r}$ is the $d$-dimensional volume element.
Then $\nabla^2 \rightarrow -\mathbf{k}^2$, yielding the ordinary differential equation (ODE):

$$\frac{d\hat{n}}{dt} = -D(t)\,\mathbf{k}^2\,\hat{n}. \tag{S5}$$

So

$$\hat{n}(\mathbf{k}, t) = \hat{n}(\mathbf{k}, 0) \exp(-\mathbf{k}^2 \Lambda(t)). \tag{S6}$$

Inverse Fourier transforming gives $n(\mathbf{r}, t)$.

3) Green's function (free space, $\mathbb{R}^d$)

The fundamental solution (point source initial condition) is the standard Gaussian kernel with $t \rightarrow \Lambda(t)$:

$$G(\mathbf{r}, t) = \frac{1}{(4\pi\Lambda(t))^{\frac{d}{2}}} \exp\left(-\frac{|\mathrm{r}|^2}{4\Lambda(t)}\right) \tag{S7}$$

For a general initial condition $n(\mathbf{r}, 0) = n_0(\mathbf{r})$,

$$n(\mathbf{r}, t) = \int d^d\mathbf{r}' \; G(\mathbf{r} - \mathbf{r}', t) \; n_0(\mathbf{r}').$$

For the 2D special case (often used for excitons in layered perovskites), the Green function is

$$G(\mathbf{r}, t) = \frac{1}{(4\pi\Lambda(t))} \exp\left(-\frac{|\mathbf{r}|^2}{4\Lambda(t)}\right). \tag{S8}$$

4) Mean-square displacement

If the total population is conserved and the distribution is localized,

$$\langle \mathbf{r}^2 \rangle = 2d\,\Lambda(t) \tag{S9}$$

Thus, in 2D: $\langle \mathbf{r}^2 \rangle = 4\Lambda(t)$, and in 3D: $\langle \mathbf{r}^2 \rangle = 6\Lambda(t)$. This corresponds to the same substitution rule, $Dt \rightarrow \Lambda(t)$, used throughout the main text (and to the relation $w^2(t) = w^2(0) + 8\Lambda(t)$ under the common optics $1/e^2$ waist convention; see Subsection 6 below).

5) Explicit $\Lambda(T)$ for the hot-carrier diffusion model

For the hot-carrier diffusion coefficient $D(t)$,

$$D(t) = D_{bulk} + \delta D\, e^{-t/t_{\mathrm{cool}}}, \tag{S10}$$

Then

$$\Lambda(t) = D_{bulk}\, t + \delta D\, t_{\mathrm{cool}} \left(1 - e^{\frac{-t}{t_{\mathrm{cool}}}}\right) \tag{S11}$$

reproducing Eq. (2) in the main text. Thus, any constant-diffusion solution can be transformed into a time-dependent diffusion solution by substituting $Dt \mapsto \Lambda(t)$.

6) Gaussian initial profile in pump–probe measurements

For an initial Gaussian exciton density profile in 2D, using the standard "1/e² waist" convention:

$$n(\mathbf{r},0) = n_0 \exp\left(-\frac{2\mathbf{r}^2}{w_0^2}\right), \tag{S12}$$

the solution remains Gaussian

$$w^2(t) = w_0^2 + 8\Lambda(t) \tag{S13}$$

$$n(\mathbf{r},t) = n_0 \left(\frac{w_0^2}{w_0^2 + 8\Lambda(t)}\right) \exp\left(-\frac{2\mathbf{r}^2}{w_0^2 + 8\Lambda(t)}\right) \tag{S14}$$

up to an overall normalization constant. This Gaussian expansion underlies the spatial overlap integrals between the Gaussian pump and the vortex probe beams that define the V-TAM signal.

7) Full V-TAM signal expression used for model fitting

Combining the time-dependent Gaussian solution derived above with the static expression for the V-TAM signal (Eq. (1) of the main text) by the substitution $D{\cdot}T \rightarrow \Lambda(T)$, the spatial-overlap contribution to the V-TAM signal becomes

$$\begin{aligned} S\big(l_{pr}, \Lambda(T), T\big) &= \frac{2}{\pi} \frac{\left(w_{pu}^2 + 8\Lambda(T)\right)^{|l_{\mathrm{pr}}|}}{\left(w_{pu}^2 + w_{pr}^2 + 8\Lambda(T)\right)^{|l_{\mathrm{pr}}|+1}} \\ &= \frac{2}{\pi} \frac{\left(w_{pu}^2 + 8(D_{bulk} \cdot T + \delta D \cdot t_{cool} \cdot (1 - e^{-T/t_{cool}}))\right)^{|l_{\mathrm{pr}}|}}{\left(w_{pu}^2 + w_{pr}^2 + 8(D_{bulk} \cdot T + \delta D \cdot t_{cool} \cdot (1 - e^{-T/t_{cool}}))\right)^{|l_{\mathrm{pr}}|+1}} \end{aligned} \tag{S15}$$

with $\Lambda(T)$ given by Eq. (2) of the main text.

The complete V-TAM signal additionally includes a mode-dependent amplitude prefactor $a_{l_{pr}}$, which accounts for differences in intensity normalization and detection efficiency across vortex modes, and an exciton population-decay term modeled as a single exponential, $e^{-T/t_{ex}}$:

$$S_{V-TAM}\big(l_{pr}, T\big) = a_{l_{pr}} \cdot S\big(l_{pr}, \Lambda(T), T\big) \cdot e^{-T/t_{ex}} \tag{S16}$$

The signal ratio with respect to the Gaussian probe is therefore

$$\begin{aligned} R_{l_{pr}0}\big(l_{pr}, T\big) &\equiv S_{V-TAM}\big(l_{pr}, T\big) / S_{V-TAM}\big(l_{pr} = 0, T\big) \\ &= \left(a_{l_{pr}}/a_0\right) \cdot S\big(l_{pr}, \Lambda(T), T\big) / S\big(0, \Lambda(T), T\big) \end{aligned} \tag{S17}$$

in which the population-decay factor $e^{-T/t_{ex}}$ cancels exactly between numerator and denominator. The signal ratios analyzed in the main text correspond to

$$R_{10}(T) \equiv S_{V-TAM}\big(l_{pr} = 1, T\big) / S_{V-TAM}\big(l_{pr} = 0, T\big) \; when \; TC = 1 \tag{S18}$$

$$R_{20}(T) \equiv S_{V-TAM}\big(l_{pr} = 2, T\big) / S_{V-TAM}\big(l_{pr} = 0, T\big) \; when \; TC = 2 \tag{S19}$$

and these are the experimentally measured quantities used to extract $D_{bulk}$, $\delta D$, and $t_{cool}$ from the fitting procedure described in the main text.

**Figure S2. IRF-based deconvolution of V-TAM spectra (520 nm)**

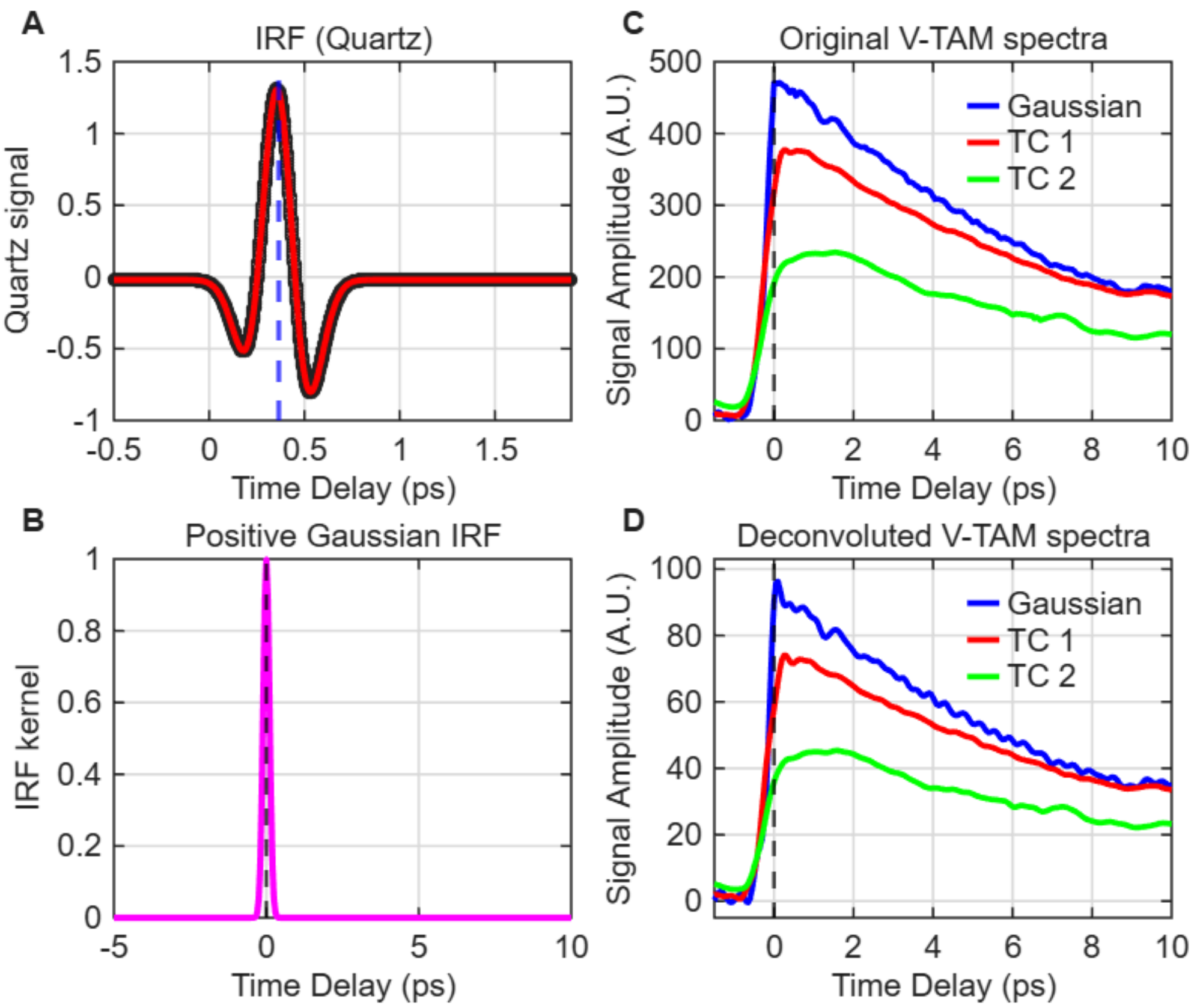

(A) The IRF was experimentally obtained from the transient signal of a quartz reference. (B) A Gaussian function was fitted to the measured IRF and recentered such that the peak position corresponds to t=0 ps, which was subsequently used as the deconvolution kernel. (C) Original V-TAM spectra measured using a Gaussian probe (blue) and vortex probes (520 nm) with topological charges (TC) of 1 (red) and 2 (green). (D) V-TAM spectra after deconvolution using the fitted Gaussian IRF kernel. As discussed in the main text, the time origin for the diffusion analysis was defined at the onset of the diffusion-dominated regime, excluding the initial signal build-up region.

**Figure S3. Frequency-domain filtering of coherent oscillations and phonon-mode assignment (520 nm probe)**

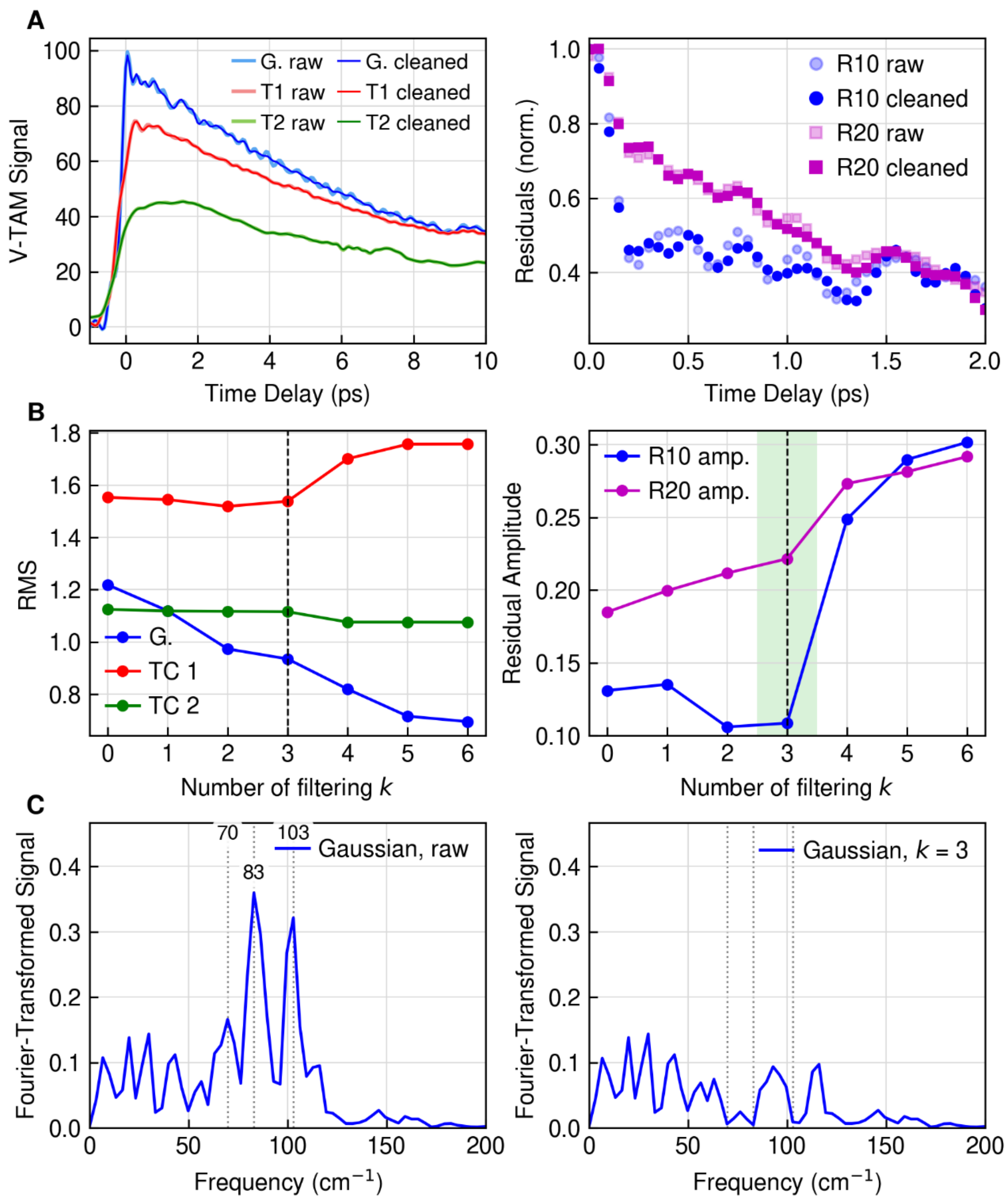

Filtered phonon modes:

| Frequency* | Mode assignment | Ref. |
|---|---|---|
| 70 $cm^{-1}$ | Pb–I bending / Pb–I–Pb scissoring | [6, 7] |
| 83 $cm^{-1}$ | Pb–I stretching | [4] |
| 103 $cm^{-1}$ | Pb–I stretching | [5] |

*Spectral resolution $\Delta\tilde{\nu} = 1/(cT) \approx 3.3$ $cm^{-1}$ for the T ≈ 10 ps analysis window.

*A. Filtering the oscillations*. Measured V-TAM signals contain periodic oscillations originating from coherent phonon modes of the $(PEA)_2PbI_4$ lattice. To isolate the underlying carrier transport dynamics from these coherent vibrational contributions, we applied an iterative Fourier-based bandstop filter to all three V-TAM channels (Gaussian, TC = 1 and 2). At each iteration, the dominant frequency in the Pb–I phonon band (25–150 cm$^{-1}$, i.e., 0.75–4.5 THz) was identified from the polynomial-detrended Fourier spectrum of the Gaussian-probe transient signal and notched out using a 4th-order Butterworth bandstop with a 0.30 THz full width. Zero-phase forward–backward filtering was used to suppress edge transients (Gibbs ringing) at the t > 9 ps endpoint of the signal.[2, 3] The identical notch set was applied to all three channels, ensuring that any common-mode oscillation cancels in the ratios $R_{10}$ and $R_{20}$. No additional smoothing of the ratios, residuals, or fits was performed. We also compare the resulting $R_{10}$ and $R_{20}$ residuals before (faint markers) and after (solid markers) filtering (right panel), showing that the oscillatory contributions are cleanly suppressed without altering the underlying decay envelope.

*B. Optimal iteration count (scree analysis).* The optimal number of iterations was determined by the scree analysis. The left panel shows the root-mean-square (RMS) amplitude of the oscillatory components as a function of filtering iteration number $k$ for each probe channel. The Gaussian RMS decreases monotonically and exhibits a clear elbow at $k = 3$, beyond which additional iterations result in only marginal suppression of the oscillation amplitude. The right panel shows the peak-to-peak amplitude of the normalized $R_{10}$ and $R_{20}$ residuals in the early-time diffusion regime. The $R_{10}$ amplitude is minimized at $k = 2$–3 (0.106 and 0.109, respectively) and then jumps sharply at $k = 4$, indicating that excessive filtering reintroduces *more* artificial oscillatory features that have the risk of distorting the intrinsic diffusion dynamics. Based on these two criteria, $k = 3$ was selected as the optimal filtering condition.

*C. Phonon mode assignment.* The frequencies filtered from the V-TAM signal fall within the well-known Pb–I lattice vibrational range (60–180 cm$^{-1}$) and can be assigned based on previous Raman and DFT studies. Panel C shows the Fourier-filtered spectrum of the Gaussian channel as a representative example, since it has the highest signal-to-noise ratio and most clearly resolves the Pb–I phonon peaks; the identical filter (at the same three frequencies) was applied to TC 1 and TC 2 channels as described in section A. The 83 cm$^{-1}$ peak corresponds to a Pb–I stretching mode, in close agreement with the calculated modes at 78.5 and 84.9 cm$^{-1}$ reported for $(PEA)_2PbI_4$ by Chen et al.[4] The 103 cm$^{-1}$ peak is also assigned to the Pb–I stretching mode, consistent with the Raman analysis of a layered hybrid perovskite by Spirito et al.[5] The 70 cm$^{-1}$ peak lies within the Pb–I–Pb bending region, characterized as the bending-related lattice modes by Pérez-Osorio et al.[6, 7] The agreement of all extracted frequencies with established lattice modes verifies that the filtered oscillatory components originate from coherent Pb–I phonons rather than detector noise or numerical artifacts, supporting the physical validity of our filtering procedure.

The T ≈ 10 ps analysis window yields a spectral resolution of $\Delta\tilde{\nu} = 1/(cT) \approx 3.3$ cm$^{-1}$. Since the three modes are separated by 13–20 cm$^{-1}$ (~4–6 bins), which is well above the discrimination limit, they are clearly resolved as distinct spectral components.

**Table S1. Optimized fitting parameters obtained from the time-dependent diffusion model analysis of the V-TAM signal ratios**

| Parameter | Value | Description |
|---|---|---|
| $w_{pu}$ | 0.196 $\mu m$ | (Gaussian) beam waist of a pump (400 nm) |
| $w_{pr}$ | 0.255 $\mu m$ | (Gaussian) beam waist of a probe (520 nm) |
| $D_{bulk}$ | 1.849 $cm^2/s$ | Steady-state (cooled) diffusion coefficient |
| $D_{initial}$ $(= \delta D)$ | 68.837 $cm^2/s$ | Initial hot-carrier diffusion enhancement |
| $t_{cool}$ | 0.347 ps | Cooling timescale |
| $a_1/a_0$ | 1.796 | Relative amplitude factor for TC = 1 |
| $a_2/a_0$ | 2.073 | Relative amplitude factor for TC = 2 |
| $R_{10}$ | 0.666 | Model initial intensity ($T = 0$) |
| $R_{20}$ | 0.366 | Model initial intensity ($T = 0$) |
| $\tau_{R10}$ | 0.250 | Cooling time from R10* |
| $\tau_{R20}$ | 0.460 | Cooling time from R20* |

* Cooling times from independent single-ratio fits.

**Figure S4. IRF-based deconvolution of V-TAM spectra (530 nm)**

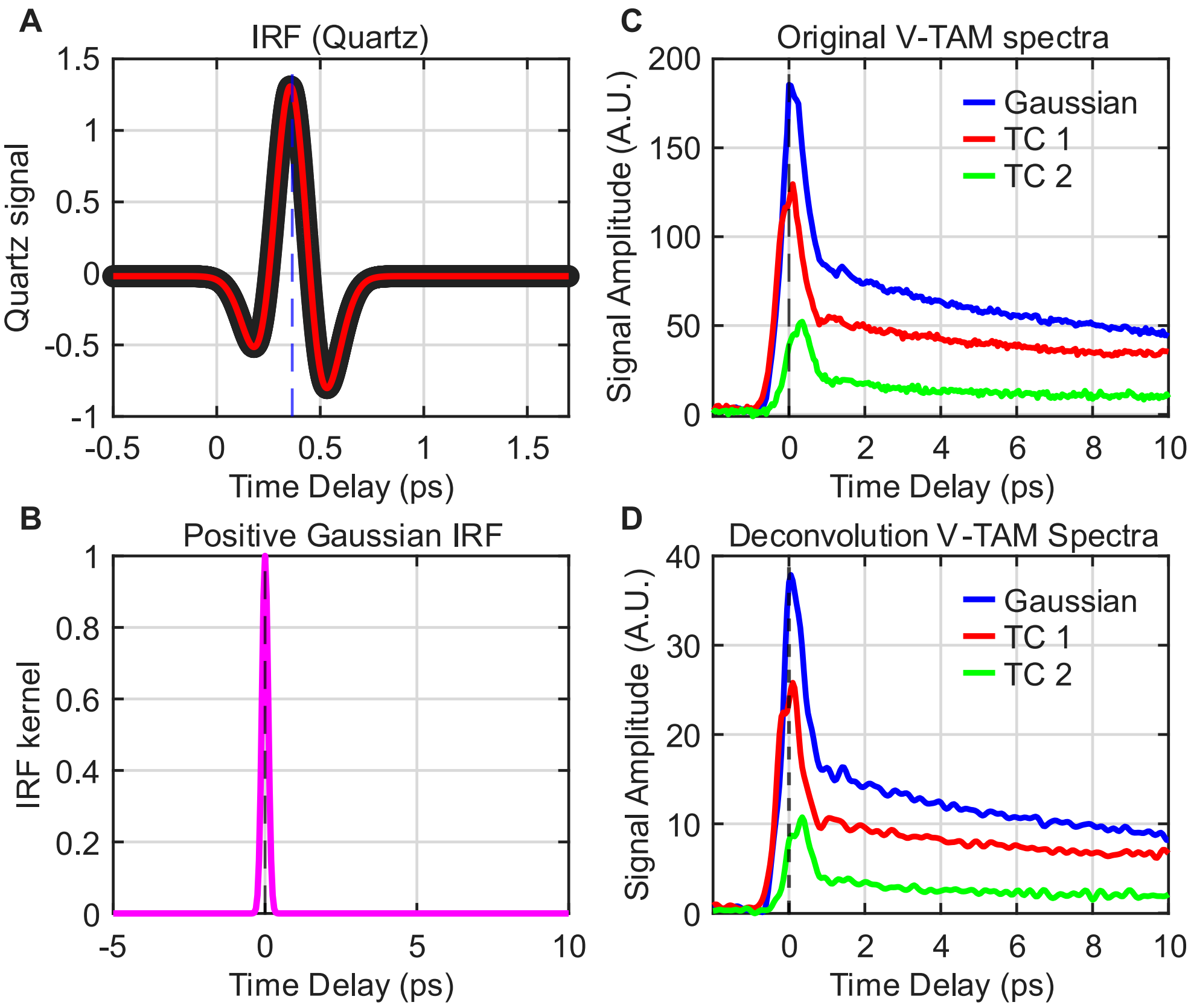

(A) The IRF was experimentally obtained from the transient signal of a quartz reference. (B) A Gaussian function was fitted to the measured IRF and recentered such that the peak position corresponds to $t = 0$ ps, which was subsequently used as the deconvolution kernel. (C) Original V-TAM spectra measured using a Gaussian probe (blue) and vortex probes (530 nm) with topological charges (TC) of 1 (red) and 2 (green). (D) V-TAM spectra after deconvolution using the fitted Gaussian IRF kernel. As discussed in the main text, the time origin for the diffusion analysis was defined at the onset of the diffusion-dominated regime, excluding the initial signal build-up region.

**Figure S5. Frequency-domain filtering of coherent oscillations and phonon-mode assignment (530 nm probe)**

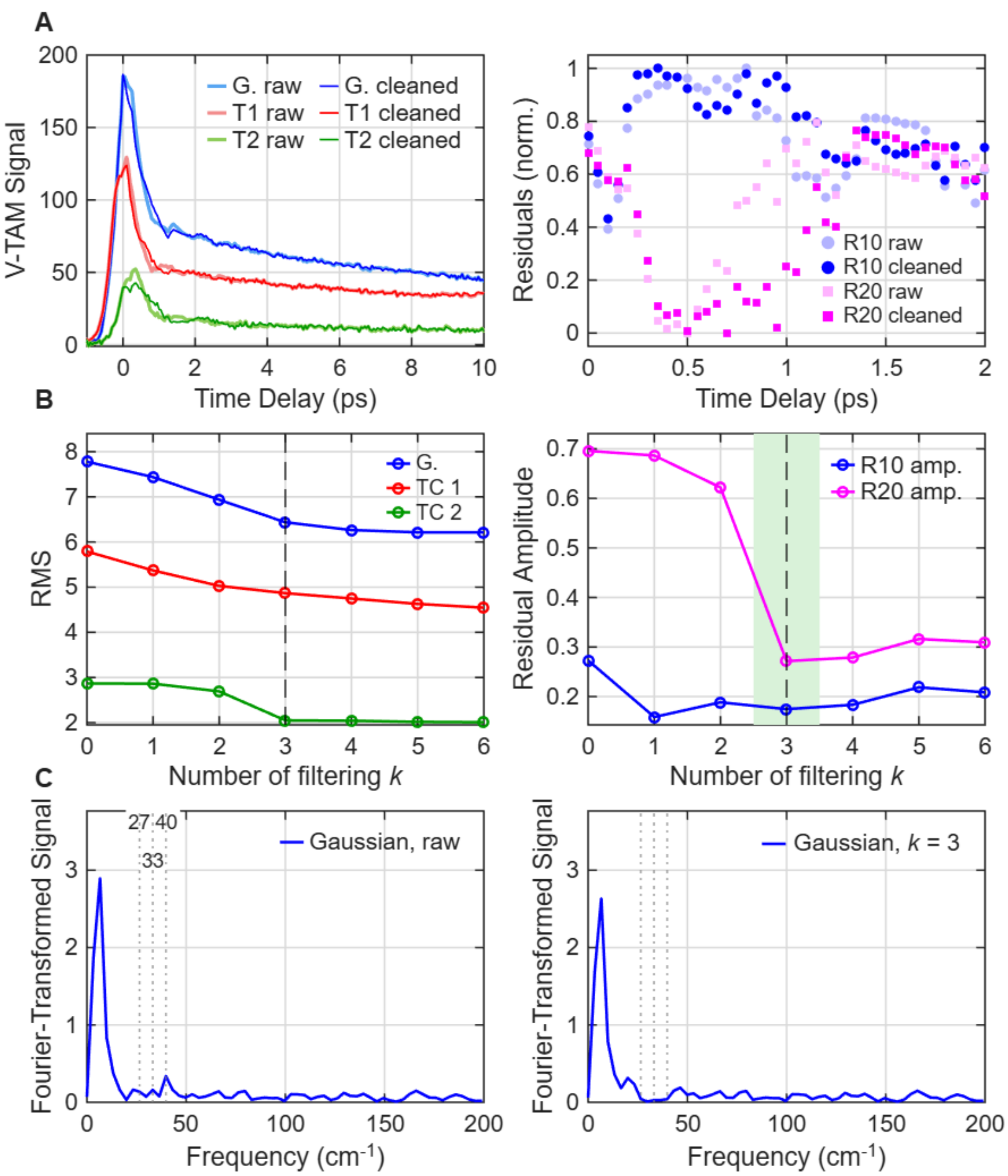

Filtered phonon modes:

| Frequency* | Mode assignment | Ref. |
|---|---|---|
| 27 $cm^{-1}$ | Zone-folded acoustic-like mode/interlayer rocking | [5] |
| 33 $cm^{-1}$ | In-plane octahedral twist + Pb–I–Pb bending | [7] |
| 40 $cm^{-1}$ | Pb–I–Pb bending (Pb-I octahedral tilt) | [5, 6] |

*Spectral resolution $\Delta\tilde{\nu} = 1/(cT) \approx 3.3$ $cm^{-1}$ for the T ≈ 10 ps analysis window.

*A. Filtering the oscillations*. The 530 nm probe V-TAM signals are also isolated from superimposed periodic oscillations due to coherent phonon modes of the perovskite lattice. The filtering procedure is identical to that used for the 520 nm probe data (Figure S3). Fourier-filtered V-TAM signals of all three channels (Gaussian, TC = 1 and 2) and the resulting ratios $R_{10}$ and $R_{20}$ before (faint markers) and after (solid markers) filtering are presented in the left and right panels, respectively.

*B. Optimal iteration count (scree analysis).* The optimal number of iterations for the 530 nm-probed data was also determined using the same scree analysis criteria described for the 520-nm probed data (Figure S3). The Gaussian RMS and the normalized $R_{10}$ and $R_{20}$ residual amplitudes simultaneously show the minimum values at $k = 3$ iterations. Additional filtering ($k \geq 4$) can reintroduce artificial oscillations, potentially distorting the intrinsic transport dynamics. Therefore, $k = 3$ was selected as the optimal filtering condition.

*C. Phonon mode assignment.* The frequencies filtered from the 530 nm V-TAM signal (27, 33, and 40 $cm^{-1}$) fall within the characteristic low-frequency Pb–I lattice vibrational range (20–60 $cm^{-1}$) and can be assigned based on previous studies. Panel C shows the Fourier-filtered spectrum of the Gaussian channel as a representative example, while the same filter was applied to TC 1 and TC 2 channels, as described in section A.

The 33 $cm^{-1}$ peak coincides with the in-plane octahedral-twist/Pb–I–Pb bending mode previously identified in $(PEA)_2PbI_4$ at near 35 $cm^{-1}$ by Rojas-Gatjens et al., one of the dominant low-energy polaronic dressing modes in this material.[7] The 40 $cm^{-1}$ peak lies within the Pb–I–Pb bending region reported for lead-iodide perovskites by Pérez-Osorio et al.[6] and Spirito et al.[5] The 27 $cm^{-1}$ is attributed to a low-frequency interlayer rocking mode or zone-folded acoustic-like lattice vibration.[5] The agreement of all extracted frequencies with established low-frequency lattice modes supports the physical validity of our filtering procedure, confirming that the removed oscillatory components originate from coherent Pb–I phonons.

Similar to Figure S3, the $T \approx 10$ ps analysis window yields a spectral resolution of $\Delta\tilde{\nu} = 1/(cT) \approx 3.3$ $cm^{-1}$. Since the three modes are separated by 13–20 $cm^{-1}$ (~4–6 bins)—well above the discrimination limit—they are clearly resolved as distinct spectral components.

**Supplementary Note 2. Increase in carrier temperature from the excess pump energy**

Given our experimental conditions:

Pump photon energy: 400 nm = 3.10 eV

(PEA)2PbI4 FE bandgap: 2.38 eV

The excess energy is therefore $\delta E =$ 3.10 - 2.38 = 0.72 eV.

Assuming the excess energy is equally partitioned between the electron and hole, each carrier acquires 0.36 eV of kinetic energy.

(a) 3D equipartition

$$T_c^{3D} = \frac{2\langle E_{kin}\rangle}{3k_B} = \frac{2 \times 0.36\ \text{eV}}{3 \times 8.617 \times 10^{-5}\text{eV/K}} \approx 2{,}780\ \text{K}$$

(b) 2D equipartition

$$T_c^{2D} = \frac{\langle E_{kin}\rangle}{k_B} = \frac{0.36\ \text{eV}}{8.617 \times 10^{-5}\text{eV/K}} \approx 4{,}180\ \text{K}$$

where $k_B = 8.617 \times 10^{-5}\text{eV/K}$ is the Boltzmann constant expressed in eV/K.

Therefore, based on the estimated excess energy from the pump in our setup, the carrier temperature $T_c$ can increase to approximately 2800–4200 K.

The enhancement in the diffusion coefficient ratio,

$$D_{hot}/D_{bulk} = (D_{bulk} + \delta D)/D_{bulk} = \frac{68.84+1.85}{1.85} = 38.21,$$

is attributed to the increase in $T_c$, along with the enhanced effective mobility $\mu$.

**Supplementary Note 3. Diffusion coefficient derived from wavelength-integrated V-TAM signal**

In the main research, the V-TAM signal was analyzed using a single probe wavelength selected with a 520 nm-centered band-pass filter, yielding an initial diffusion coefficient $D_{initial} =$ 68.84 $cm^2$/s.

Instead, if a wavelength-integrated signal over the 500–580 nm range is used for the V-TAM analysis, the extracted $D_{initial}$ value changes significantly.

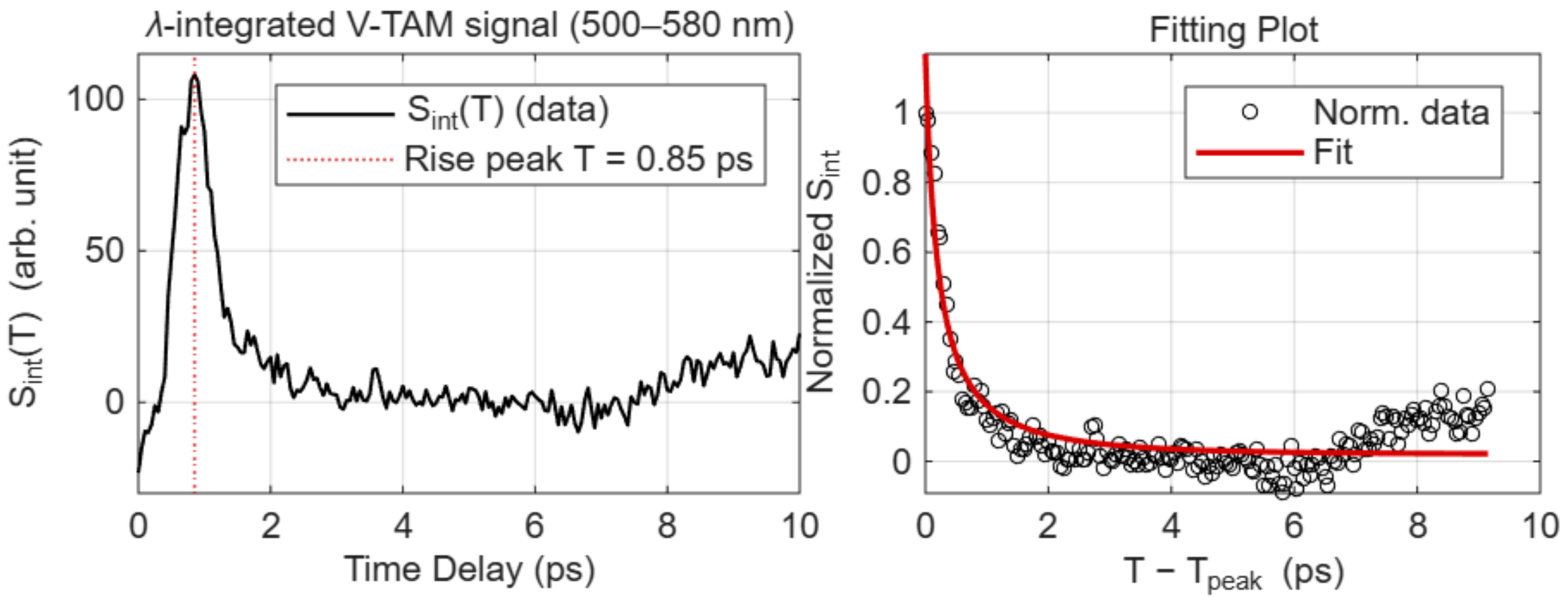

The left panel shows the V-TAM signal integrated over the 500–580 nm spectral range. By fitting the signal using the spatial diffusion model (Eq. (1) in the main text) with a Gaussian probe ($l_{pr} = 0$) (right panel), the extracted diffusion coefficient is $D = 15.4 \pm 22.7 \text{cm}^2/\text{s}$, which is approximately 4-5 times smaller than the value obtained from the single-wavelength probe measurement.

This result demonstrates that the extracted diffusion coefficient is highly sensitive to the spectral range of the probe used in the analysis. When a broad wavelength-integrated signal is employed, contributions from carriers with different spectral and dynamical characteristics are averaged, leading to a substantially underestimated diffusion coefficient and large fitting uncertainty. In contrast, the single-wavelength (narrow-band) probe selectively tracks a well-defined carrier population, enabling more reliable extraction of the intrinsic ultrafast diffusion dynamics. These results highlight the importance of spectrally selective probing for accurately resolving nonequilibrium carrier transport in ultrafast TAM measurements.

**Figure S6. Schematic illustration of V-TAM optical setup**

To investigate the ultrafast diffusion dynamics of free excitons and exciton-polarons, a custom-built TAM system was developed for transmission-mode operation. Fundamental seed pulses (100-fs pulse width, 1030-nm wavelength) were generated using a Light Conversion Yb:KGW laser system (PHAROS), delivering a repetition rate of 50 kHz and an average output power of 6 W. The laser output was split into two arms for pump and probe pulse generation.

In the pump arm, the beam was directed into an optical parametric amplifier (OPA; Orpheus) to generate 800 nm pulses. After passing through a mechanical chopper, the beam was frequency-doubled in a beta-barium borate (BBO) crystal, generating 400 nm pulses. A 400 nm bandpass filter was used to isolate the pump beam while suppressing residual 800 nm light. In the probe arm, a portion of the fundamental beam was focused into a YAG crystal to generate a white-light continuum. The continuum probe was spectrally filtered using a bandpass filter centered at 520 or 530 nm with a 10 nm spectral width (Edmund Optics), enabling single-color pump-probe measurements of free-exciton or exciton-polaron dynamics, respectively. The filtered probe beam was then passed through an optical vortex plate (TC =1 and 2, Thorlabs) to generate a donut-shaped probe beam. The laser intensity is controlled by a half-wave plate (λ/2) and a polarizing beam splitter (PBS), followed by beam expansion using a two-lens system (focal lengths of 100 and 50 mm). The pump and probe beams were collinearly combined and focused onto the sample using an objective lens (numerical aperture, NA = 0.65), yielding tightly focused beams with diameters of 0.1959 μm for the pump, and 0.2546 μm and 0.2596 μm for the 520 and 530 nm probe, respectively. The transmitted probe signal was collected using an optical fiber and analyzed with an Ocean Insight spectrometer (OCEAN FX). To suppress intense pump scattering, a 455 nm long-pass filter was placed before the optical fiber. The spatial overlap between the pump and probe beams was verified using a CMOS camera (Thorlabs). The instrumental response function of the TAM setup was determined using a quartz plate with an intense pump pulse.

## Supplementary references